\documentstyle[12pt,pictex]{ioplppt}
\begin{document}
\jl{1}
\title{Asymptotic form of the approach to equilibrium in reversible
recombination reactions}[Approach to equilibrium in reversible
reactions]
\author{Pierre-Antoine Rey\dag\ and John Cardy\dag\ddag}
\address{\dag Theoretical Physics, Oxford University, 1 Keble road, Oxford
OX1 3NP, United Kingdom}
\address{\ddag All Souls College, Oxford}

\begin{abstract}
The reversible reactions $A+A\rightleftharpoons C$ and
$A+B\rightleftharpoons C$ are investigated. From the exact Langevin
equations describing our model, we set up a systematic approximation
scheme to compute the approach of the density of $C$ particles to its
equilibrium value. We show that for sufficiently long time $t$, this
approach takes the form of a power law $At^{-d/2}$, for any dimension
$d$. The amplitude $A$ is also computed exactly, but is expected to be
model dependent. For uncorrelated initial conditions, the $C$ density
turns out to be a monotonic time function. The cases of correlated
initial conditions and unequal diffusion constants are investigated as
well. In the former, correlations may break the monotonicity of the
density or in some special cases they may change the long time
behavior. For the latter, the power law remains valid, only the
amplitude changes, even in the extreme case of immobile $C$
particles. We also consider the case of segregated initial condition
for which a reaction front is observed, and confirm that its width is
governed by mean-field exponent in any dimension.
\end{abstract}

\maketitle

\section{Introduction}

During the last two decades, diffusion-limited chemical reactions have
attracted considerable interest. In particular, the one- and
two-species annihilation reactions $A+A\to C$ and $A+B\to C$ are known
to exhibit anomalous kinetics in lower dimension. For the one-species
case, the upper critical dimension --- the dimension above which the
rate equation $\dot n_a(t) = -k n_a^2(t)$ (where $n_a(t)$ is the
concentration of $A$ particles) gives qualitatively correct results
--- is 2 \cite{AA1,AA2,AA3,Lee,RD}. For the two-species case, the
situation is more complex, because of the presence of a conserved
quantity, namely the difference concentration of $A$ and $B$
particles. Depending on the initial conditions, the upper-critical
dimension is 4 (for homogeneous conditions with the same initial
concentration of both particle types) or 2 (for segregated initial
conditions or non-equal initial $A$ and $B$ concentration, etc.)
\cite{AA1,AB1,AB2,LC,HC}. Among other methods, those results have been
derived using renormalization group techniques, drawing a rather
complete picture of the different universality classes involved in
these reactions \cite{Lee,RD,LC,HC}.

In this paper, the question we want to address is the following: what
happens if the backward decombination reaction is allowed with a
given probability? In most physically interesting systems it is
unlikely that this possibility is totally forbidden. Although for
extremely small backward probability, one expects the effects to be
very small (and even unnoticeable for not too large observation time),
a fundamental change occurs in the system. Instead of decreasing
toward a non-equilibrium steady-state, the system should eventually
reach an equilibrium state. Moreover a new conserved quantity can be
constructed reflecting the conservation of mass or energy:
$n_a(t)+2n_c(t)$ for the one-species reversible reaction
$A+A\rightleftharpoons C$ and $n_a(t)+n_b(t)+2n_c(t)$ for the two-species
reaction $A+B\rightleftharpoons C$ (where $n_b(t)$ and $n_c(t)$ are
respectively the $B$ and $C$ particles concentration).

At the mean-field level, the rate equations give an exponential
approach toward the equilibrium state. However it was soon recognized
\cite{ZO1,ZO2,KR,BOO} that the conserved quantity obeys a diffusion
equation, and thus its initial fluctuations should decay with a power
law: $t^{-d/2}$ ($d$ is the dimension of the system). In the long time
limit, this power will always overcome the exponentially fast decay of
the rate equations, for {\it any} dimension $d>0$. The upper critical
dimension is thus infinity. At first sight surprising, this result has
been confirmed in various ways. For the single-species reaction,
Zeldovich and Ovchinnikov \cite{ZO2} obtained the approach to
equilibrium in the low density limit of a field theory in three
dimensions. This result was extended later \cite{BOO} for arbitrary
dimensions, but in the framework of an uncontrolled approximation. In
the $A+B\rightleftharpoons C$ case, the power law decay was first
obtained in three dimensions \cite{ZO1} from heuristic
considerations. Later Kang and Redner \cite{KR}, using an argument
based on the fluctuations, and Burlastky \etal \cite{BOO}, assuming a
closure of the hierarchy, extended this result to arbitrary
dimensions, but with different amplitudes. Our purpose in this paper is
to derive, using field theory techniques and Langevin equations, the
asymptotic approach to equilibrium in a controlled and exact
way. Whereas we are able to show the universality of the power law
exponent, the amplitude proves to be model-dependent.

The paper is organized as follows. In \sref{sec:AAC}, we consider the
single-species $A+A\rightleftharpoons C$ reaction. From the master
equation describing our model, we map the problem to a set of two
Langevin equations for the random variables $a$ and $c$, with complex
noise. The concentrations $n_a(t)$ and $n_c(t)$ are then given by the
average of the random variables $a$ and $c$ over the noise:
$n_a=\langle a \rangle$ and $n_c=\langle c \rangle$. Exploiting the
fact that $n_a(t)+2n_c(t)$ is conserved by the dynamics, i.e.\ that
$a+2c$ obeys a noisy diffusion equation, we can write down the long
time behavior of its two-point correlation:
\begin{equation}
\langle(a+2c)^2\rangle - \langle a+2c \rangle^2
\sim (c_\infty -c_0)(8\pi Dt)^{-d/2}
\end{equation}
($c_\infty$ is the steady-state density of $C$ particles, $c_0$ the
initial one, and $D$ the diffusion constant of both $A$ and $C$
particles). We can then set up a controlled approximation scheme to
obtain the approach to the equilibrium, based on the fact that the
previous correlation goes to zero as $t$ grows to infinity. We find
that
\begin{equation}
\langle c \rangle - c_\infty
\sim \frac{2\lambda\mu^2}{(4\lambda a_\infty + \mu)^3}(c_0-c_\infty)
    (8\pi Dt)^{-d/2}
\end{equation}
($\lambda$ and $\mu$ are respectively the rate of the forward and
backward reactions and $a_\infty$ is the equilibrium density of $A$
particles). The exponent of the power law is universal, as it comes
from the conservation law, but the amplitude is model-dependent. It is
however interesting to note that Burlatsky \etal \cite{BOO} found the
same result (up to a factor of 2) for a slightly different
model, their results relying, nevertheless, on an uncontrolled
approximation (the closure of the hierarchy). Note that the final
approach to equilibrium is governed by the sign of $c_0-c_\infty$. In
fact, the density is expected to reach its equilibrium value
monotonically.

In \sref{sec:ABC}, we consider the reaction $A+B\rightleftharpoons
C$. Using the same method (in this case there is a second quantity for
which the Langevin equation can also be solved exactly), we find
\begin{equation}
\langle c \rangle - c_\infty
\sim \frac{\lambda}{2\sigma_{\rm\!A\!B}^3}
     [\sigma_{\rm\!A\!B}^2 + \mu^2 - \lambda^2(a_0-b_0)^2]
     (c_0-c_\infty)(8\pi Dt)^{-d/2}
\end{equation}
where $\sigma_{\rm\!A\!B}=\lambda(a_0+b_0+2c_0-2c_\infty)+\mu$, and
$a_0$ and $b_0$ are the initial $A$ and $B$ particle densities. An
equivalent expression were obtained by Burlatsky \etal \cite{BOO} when
$a_0=b_0$. Similar conclusions concerning universality and
monotonicity can be drawn.

In \sref{sec:ext}, various generalizations are considered. First we
investigate the case of pair correlations between the reactants
(i.e. $A$-$A$ pairs for $A+A\rightleftharpoons C$ and $A$-$B$ pairs
for $A+B\rightleftharpoons C$). We show that, depending on the initial
fraction of correlated particles, the approach to equilibrium can
become non-monotonic, or even, in some special cases, the power law
may change to a faster decay. The second part of this section is
devoted to the interesting problem of unequal diffusion constants for
the different species. Even in extreme cases (such as immobile $C$
particles), the power law is unchanged and only the amplitude is
modified. The last generalization we consider is the case of
segregated initial conditions. When initially the $A$ and $B$
particles are spatially separated, a reaction front will develop. One
natural question to ask is to which rapidity the width of this front
will grow. We are able to confirm the scaling results obtained by
Chopard \etal \cite{CDKR} showing that this increase is governed by
the mean-field behavior $w(t)\sim t^{1/2}$. This result is also valid
for immobile $C$ particles. Final remarks are made in \sref{sec:conc}.

\section{The $A+A\protect\rightleftharpoons C$ reaction}
\label{sec:AAC}

\subsection{The model and the formalism used}

Our starting point is the following continuous time master equation
\begin{eqnarray}\fl
\frac{\partial}{\partial t} P(\{m\},\{n\};t) \nonumber \\
\lo=\frac{D_a}{\ell^2}\sum_i\sum_{e_i}
    [(m_{e_i}+1)P(\ldots,m_i-1,m_{e_i}+1,\ldots,\{n\};t) \nonumber \\
{}-  m_iP(\{m\},\{n\};t)] \nonumber \\
{}+ \frac{D_c}{\ell^2}\sum_i\sum_{e_i}
    [(n_{e_i}+1)P(\{m\},\ldots,n_i-1,n_{e_i}+1,\ldots;t) \nonumber \\
{}-  n_iP(\{m\},\{n\};t)] \nonumber \\
{}+ \lambda_0\sum_i
    [(m_i+2)(m_i+1)P(\ldots,m_i+2,\ldots,\{n\};t) \nonumber \\
{}-  m_i(m_i-1)P(\{m\},\{n\};t)] \nonumber \\
{}+ \mu\sum_i[(n_i+1)P(\{m\},\ldots,n_i+2,\ldots;t)
  - n_iP(\{m\},\{n\};t)].
\label{eq:ME}
\end{eqnarray}
The set $\{m\}$ ($\{n\}$) denotes the occupation numbers of $A$ ($C$)
particles in each lattice site and $P(\{m\},\{n\};t)$ is the
probability to find  the configuration $\{m\}$, $\{n\}$ at time
$t$. This equation describes the evolution of the probability $P$ in
time. A given configuration can change due to four processes: by
diffusion of $A$ particles (second and third lines of \eref{eq:ME},
where $D_a$ is the diffusion constant of the $A$ particles and $\ell$
is the lattice constant); by diffusion of $C$ particles (with
diffusion constant $D_c$, fourth and fifth lines). It will also change
when two $A$ particles merge into one $C$, with a microscopic reaction
rate $\lambda_0$ (sixth and seventh lines) or, as shown in the last
line of \eref{eq:ME}, when a $C$ particle reacts producing two $A$
(the corresponding rate is denoted by $\mu$). In the diffusion terms,
the sum over $e_i$ stands for a summation over all the nearest
neighbors of site $i$. In this respect, equation \eref{eq:ME} models
then the time continuous evolution of the reversible reaction
$A+A\rightleftharpoons C$ on a $d$-dimensional hypercubic lattice,
allowing for multiple occupancy on each site.

For the time being, we choose the initial conditions to be given by an
uncorrelated Poissonian distribution on each site and for each
species:
\begin{equation}
P(\{m\},\{n\};0)
  = \e^{-\tilde a_0 - \tilde c_0}
    \prod_i \frac{\tilde a_0^{m_i}}{m_i!}
    \frac{\tilde c_0^{n_i}}{n_i!},
\label{eq:CI}
\end{equation}
where $\tilde a_0$ ($\tilde c_0$) is the average occupation number per
lattice site for the $A$ ($C$) particles. A correlated initial condition
will be considered later, in \sref{sec:ext}.

More than twenty years ago, Doi \cite{Doi} (see also \cite{GS})
developed a procedure mapping the master equation to a second
quantized representation by introducing sets of creation and
annihilation operators. In turn this second quantized form can be
mapped to a field theory (see \cite{Pel}). By now these various steps
are well known and we shall only quote the results we need.

Let us introduce the (complex) fields $a$, $\bar a$, $c$ and $\bar
c$. In terms of these fields, the $A$ particle density is given by
$n_a({\bf x},t)=\langle\!\langle a({\bf x},t) \rangle\!\rangle$ where
$\langle\!\langle \cdot \rangle\!\rangle$ denotes an average over
$\e^{-S}$, where $S$ is an action. In general,
\begin{equation}
\langle\!\langle {\cal A}[a,c] \rangle\!\rangle
  = \int\!{\cal D}a\,{\cal D}\bar a\,{\cal D}c\,{\cal D}\bar c\,
    {\cal A}[a,c]\,\e^{-S}
\bigg /\!
    \int\!{\cal D}a\,{\cal D}\bar a\,{\cal D}c\,{\cal D}\bar c\,\e^{-S}.
    \label{eq:moy}
\end{equation}
The script $\cal D$ denotes functional integration and $S$ is the
action corresponding to our reaction, obtained by the mapping of the
master equation \eref{eq:ME}
\begin{eqnarray}\fl
S = \int\!\d^dx\!\int_0^{t_{\rm f}}\d t\,
    [\bar a(\partial_t - D_a\nabla^2)a + \bar c(\partial_t - D_c\nabla^2)c
  +  (\bar a^2 + 2\bar a - \bar c)(\lambda a^2 - \mu c) \nonumber \\
{}-  \delta(t)(a_0\bar a + c_0\bar c)] \label{eq:act},
\end{eqnarray}
where $\lambda=\lambda_0\ell^d$, $a_0=\tilde a_0/\ell^d$ and
$c_0=\tilde c_0/\ell^d$. For convenience, the continuous space limit
($\ell\to0$) has been taken. However an equivalent result can be
written keeping the lattice structure. Note that any observable $\cal
A$ can always be written as an expression which solely depends on the
fields $a$ and $c$ and not on the response fields $\bar a$ and $\bar
c$ (this comes from probability conservation). The double bracket
notation in \eref{eq:moy} stresses the fact that average is taken both
over the dynamics (first line of \eref{eq:act}) and the initial
conditions (second line).

The density correlation function for the $A$ particles is given by
\begin{equation}
C_a({\bf x},t)
  = \langle\!\langle[a({\bf x},t)+\delta^d({\bf
x})]a(0,t)\rangle\!\rangle.
\label{eq:corrfct}
\end{equation}
Similar relations hold for the $C$ particles.

The analysis of the action using the renormalization group formalism has
prowed to be extremely powerful for various types of reaction (see
\cite{Lee,RD,LC,HC} for some recent examples). In particular it is
especially well suited to distinguish universality classes. However, in
our case, this analysis is not well adapted because the upper critical
dimension is infinite and an expansion around it will clearly
fail. In fact, to treat this problem, we shall prefer to use the
formalism of Langevin equations. These equations can be very easily
obtained by replacing the quartic piece of the action with an integral
over a noise variable:
\begin{equation}
\exp[-\bar a^2(\lambda a^2 - \mu c)]
\sim \int_{-\infty}^{+\infty}\d\zeta\,
     \exp[\bar a\zeta - \case14\zeta^2/(\mu c - \lambda a^2)]
\end{equation}
and integrating out the response fields $\bar a$ and $\bar
c$. One then obtains the two following equations
\begin{eqnarray}
(\partial_t - D_a\nabla^2) a({\bf x},t)
  = -2\lambda a({\bf x},t)^2 + 2\mu c({\bf x},t) + \zeta({\bf x},t)
    \label{eq:ALan} \\
(\partial_t - D_c\nabla^2) c({\bf x},t)
  = \lambda a({\bf x},t)^2 - \mu c({\bf x},t), \label{eq:CLan}
\end{eqnarray}
where $\zeta$ is a complex Gaussian noise with zero mean value, whose
correlation is given by
\begin{equation}
\langle \zeta({\bf x},t)\zeta({\bf x}',t') \rangle
  = 2\langle \mu c({\bf x},t) - \lambda a({\bf x},t)^2 \rangle
    \delta^{(d)}({\bf x}-{\bf x}') \delta(t-t').
\end{equation}
Here, the single bracket notation stands for the average over the
noise. There is no more need to average over the initial conditions as
it has been explicitly done when integrating over the response field.

Note that by using \eref{eq:CLan}, one has, for homogeneous initial
conditions,
\begin{equation}
\langle \zeta({\bf x},t)\zeta({\bf x}',t') \rangle
  = -2\partial_t\langle c(t)\rangle\delta^{(d)}({\bf x}-{\bf x}')
    \delta(t-t').\label{eq:corr}
\end{equation}
As the density is expected to reach a (reversible) stationary state,
the noise correlation should vanish in the long time limit
($\lim_{t\to\infty} \partial_t\langle c(t)\rangle=0$). This,
together with the fact that the equation for the variable $c({\bf
x},t)$ comes without explicit noise (except through the $a$
dependence) are the central points of our analysis.

These Langevin equations look very similar to the rate equations,
with the addition of the noise. One then might question the necessity
of deriving them in a such complicated way, as it would have been
easier just to add noise to the rate equation. In fact this method
allows us to derive exactly the noise and its noise-noise correlation
function. In particular one sees that the noise is complex, a result
which is at odds with the usual guess made when writing heuristic
Langevin equations. It should also be emphasized that the variables
$a({\bf x},t)$ and $c({\bf x},t)$ do {\em not}\/ represent the
density, because they are complex. The mean density $n_a(t)$
($n_c(t)$) is given by the average of $a({\bf x},t)$ ($c({\bf x},t)$)
over the noise. One easily convinces oneself that this average give a
real value for the density.

An immediate consequence of the vanishing of the fluctuations at
equilibrium is the possibility of computing the actual values of
the equilibrium densities, which are simply given by their mean field
values. Denoting by $a_\infty$ and $c_\infty$ such densities, one has
\begin{equation}
\lambda a_\infty^2 = \mu c_\infty, \label{eq:steady}
\end{equation}
which together with the conservation law
\begin{equation}
a_\infty + 2 c_\infty = a_0 + 2 c_0,
\end{equation}
give us
\begin{equation}
a_\infty = \frac{\mu}{4\lambda}
           \Bigl(\sqrt{1+8\lambda(a_0+2c_0)/\mu}-1\Bigr).
\end{equation}
It is easy to check that this result is a solution of the detailed
balance condition of the master equation \eref{eq:ME}, which should hold
as the stationary state is an equilibrium state.
Note that we do not expect this result to be universal,
i.e.\ apply for all models describing a reversible
$A+A\rightleftharpoons C$ reaction. In fact this result strongly
relies on the multiple occupancy property and single site reactions of
our model. It can be indeed shown \cite{ZolP} that, in one dimension,
a spin chain model of this reaction with exclusion process leads to a
different steady-state density (equation \eref{eq:steady} is replaced
by another condition). However in the small density limit (dilute gas)
this latter result converges toward the mean field value which is
expected to be universal in this regime.

\subsection{Conservation law}

The next step in our analysis of the model is, of course, to obtain the
approach toward equilibrium. It is easily seen that the rate equations
give an exponential decay. The fluctuations are expected, however, to
change this law. Our starting point to analyze this problem will be
the two equations \eref{eq:ALan} and \eref{eq:CLan} which, with the
noise-noise correlation, describe completely our model. In order to
simplify, we shall consider now the case of equal diffusion constant
$D_a=D_c\equiv D$. The case for which $D_a\neq D_c$ will be considered
in \sref{sec:ext}.

First one remarks that the quantity $\chi = a + 2c$ obeys a noisy
diffusion equation:
\begin{equation}
(\partial_t - D_a\nabla^2) \chi({\bf x},t) = \zeta({\bf x},t)
\label{eq:cons}
\end{equation}
This reflects the fact that the quantity $n_a({\bf x},t)+2n_c({\bf
x},t)=\langle\chi({\bf x},t)\rangle$ is conserved by the dynamics, which in
turn is a statement about mass (or energy)
conservation. \Eref{eq:cons} is easily solved, and one finds
\begin{equation}\fl
\chi({\bf x},t)
  = \int_0^t\d t'\!\int\!\d^dx'\,
    G_0({\bf x}-{\bf x}',t-t')\zeta({\bf x}',t')
  + \int\!\d^dx'\,G_0({\bf x}-{\bf x}',t)\chi({\bf x}',0)
    \label{eq:solchi}
\end{equation}
where $\chi({\bf x},0)$ is the initial condition and $G_0({\bf x},t)$
is the free propagator:
\begin{equation}
G_0({\bf x},t)
  = \theta(t)\,(4\pi Dt)^{-d/2}\exp\biggl(-\frac{x^2}{4Dt}\biggr)
\end{equation}
($\theta(t)$ is the usual Heaviside step function). Using
$\langle\zeta\rangle=0$,
\begin{equation}
\langle\chi({\bf x},t)^2\rangle - \langle\chi({\bf x},t)\rangle^2
  = -2\int_0^t\d t_1\,[8\pi D(t-t_1)]^{-d/2}
    \partial_t\langle c(t_1)\rangle, \label{eq:chichi}
\end{equation}
with $\langle\chi({\bf x},t)\rangle = a_0 + 2c_0$. Although the exact
structure of $\langle c(t)\rangle$ is not known (as this is precisely
the quantity we want to compute), we do not need it to obtain
$\langle\chi({\bf x},t)^2\rangle$ for long time, as, when $t\to\infty$
\begin{equation}\fl
\langle\chi({\bf x},t)^2\rangle - \langle\chi({\bf x},t)\rangle^2
  = -2 [8\pi Dt]^{-d/2}\int_0^\infty\!\d t_1\,
    \partial_t\langle c(t_1)\rangle
  = -2 (c_\infty - c_0) [8\pi Dt]^{-d/2}. \label{eq:chichilong}
\end{equation}

At this stage several remarks have to be made. First, the integrand of
expression \eref{eq:chichi} diverges when $t_1\to t$. For $d\geq2$ the
integral is thus singular. This divergence is however artificial as it
comes from the continuous space limit we took when writing the action,
and it can be avoided by putting a short distance cut-off of the order
of $\ell$ in the space integration. In turn, once the integration over
space is performed, this small distance cut-off will produce a cut-off
function ${\cal C}_{\rm cf}(\ell^2/D(t-t_1))$ which multiply the
integrand of equation \eref{eq:chichi}. The exact form of this cut-off
function is unimportant. It should be a rapidly decreasing function
for large $x$ and it should go to 1 when $x$ goes to 0 (for example a
possible candidate could be ${\cal C}_{\rm cf}(x)\sim \exp(-x^2)$). In
the following, such regularization will always been understood when
facing ultra-violet divergent integrals. A second remark concerns the
sign of the variance of $\chi$ which can be positive or negative,
depending on the initial densities $a_0$ and $c_0$, i.e.\ on the sign
of $c_0-c_\infty$. (Note that $c_\infty$ depends both on $a_0$ and
$c_0$ and it is always possible to adjust $a_0$ such that
$c_\infty<c_0$, or $c_\infty>c_0$.) The possibility of having negative
variance plays a central role. In the next subsection, we shall show
that the approach toward equilibrium is monotonic, at least for times
at which our analysis apply. Whereas this result is not really
surprising as it was already obtained by the rate equation, one should
note that it refutes the results of a common method of solving such
problems. From the two Langevin equations \eref{eq:ALan} and
\eref{eq:CLan}, a natural approximation for the density would be to
consider the standard rate equations
\begin{eqnarray}
(\partial_t - D\nabla^2)\hat a({\bf x},t)
  = -2\lambda \hat a({\bf x},t)^2 + 2\mu \hat c({\bf x},t) \\
(\partial_t - D\nabla^2)\hat c({\bf x},t)
  = \lambda \hat a({\bf x},t)^2 - \mu \hat c({\bf x},t)
\end{eqnarray}
with Poissonian random initial conditions. Whereas this approximation
gives perfectly good results, both for short and long times (but not
for intermediate times), for the two species annihilation reaction
$A+B\to\emptyset$ (see for example \cite{LC,DP}), in our case it fails to
predict the anti-correlation of the conserved field. Indeed, it is easy
to solve the equation for $\hat \chi=\hat a+2\hat c$. One has
\begin{equation}
\hat \chi({\bf x},t)
  = \int\!\d^dx'\,G_0({\bf x}-{\bf x}',t)\hat \chi({\bf x}',0).
\end{equation}
Denoting the average over the initial conditions by
$\langle\cdot\rangle_{\rm p}$, one readily finds that the density
$\langle\hat\chi\rangle_{\rm p}$ is conserved:
\begin{equation}
\langle\hat\chi({\bf x},t)\rangle_{\rm p} = a_0 + 2 c_0.
\end{equation}
However, due to the Poissonian initial conditions which imply
\begin{eqnarray}
\langle \hat a({\bf x},0)\hat a({\bf x}',0)\rangle_{\rm p}
  = a_0^2 + a_0 \delta^{(d)}({\bf x}-{\bf x}'), \\
\langle \hat c({\bf x},0)\hat c({\bf x}',0)\rangle_{\rm p}
  = c_0^2 + c_0 \delta^{(d)}({\bf x}-{\bf x}'),
\end{eqnarray}
one finds
\begin{equation}
\langle\hat \chi({\bf x},t)^2\rangle_{\rm p}
  = (a_0 + 2 c_0)^2 + (a_0 + 4 c_0)(8\pi Dt)^{-d/2}.
\end{equation}
This approximation gives satisfactory results concerning the power law
approach to equilibrium, however it is unable to predict the correct
sign of the correlations. As a consequence of this erroneous sign, the
density of the $C$ particles would always approach its stationary
value from above, leading to a non-monotonic behavior, when
$c_0<c_\infty$.

\subsection{Approximation scheme for the concentration of $C$ particles}

In this subsection, we would like to compute the approach of the density
to its stationary value. Let us define $\delta\!c({\bf x},t) = c({\bf
x},t) - c_\infty$ and $\delta\!\chi({\bf x},t) = \chi({\bf x},t) - (a_0
+ 2c_0)$. The Langevin equation for $\delta\!c$ is then given by
\begin{eqnarray}\fl
(\partial_t - D\nabla^2 + \sigma_{\rm\!A\!A})
\,\delta\!c({\bf x},t) \nonumber \\
\lo=4\lambda\,\delta\!c({\bf x},t)^2
  - 4\lambda\,\delta\!\chi({\bf x},t) \,\delta\!c({\bf x},t)
  + \lambda\,\delta\!\chi({\bf x},t)^2
  + \case12(\sigma_{\rm\!A\!A}-\mu)\,\delta\!\chi({\bf x},t)
\label{eq:deltac}
\end{eqnarray}
where we put $\sigma_{\rm\!A\!A}=4\lambda a_\infty + \mu$. The
explicit solution of \eref{eq:deltac} is unknown. However, we can
obtain the large time behavior of $\langle\delta\!c\rangle$, by
exploiting that $\delta\!\chi$ is a Gaussian random variable with
vanishing variance when $t$ goes to infinity. The formal solution of
\eref{eq:deltac} can be written in the form
\begin{equation}
\delta\!c
  = {\cal G}_0[\delta\!c]
  + {\cal G}[4\lambda\,\delta\!c^2 - 4\lambda\,\delta\!\chi\,\delta\!c
  + \lambda\,\delta\!\chi^2
  + \case12(\sigma_{\rm\!A\!A}-\mu)\,\delta\!\chi], \label{eq:formsol}
\end{equation}
where to simplify our notation we have introduced
\begin{equation}
{\cal G}[f]({\bf x},t)
  = \int_0^t\d t\!\int\d^dx'\,\e^{-\sigma_{\rm\!A\!A}(t-t')}
    G_0({\bf x}-{\bf x}',t-t') f({\bf x}',t') \label{eq:defG}
\end{equation}
and
\begin{equation}
{\cal G}_0[f]({\bf x},t)
  = \int\d^dx'\,\e^{-\sigma_{\rm\!A\!A}t}G_0({\bf x}-{\bf x}',t)
    f({\bf x}',0). \label{eq:defG0}
\end{equation}
The first term in \eref{eq:formsol} comes from the initial
condition. It simply reduces to
$(c_0-c_\infty)\e^{-\sigma_{\rm\!A\!A}t}$.

Iterating this solution will give eventually a series in ${\cal
G}_0[\delta\!c]$, ${\cal G}[\delta\!\chi]$ and ${\cal
G}[\delta\!\chi^2]$, with appropriate insertions of the operator $\cal
G$. Three kinds of terms then occur: terms containing only power of
${\cal G}_0[\delta\!c]$, terms containing only power of ${\cal
G}[\delta\!\chi]$ and ${\cal G}[\delta\!\chi^2]$, and mixed terms. One
easily sees that the first kind of terms gives exponential decay in the
long time limit, they can thus be discarded as we are interested in
the asymptotic time regime of $\langle\delta\!c\rangle$ and
$\langle\delta\!c^2\rangle$. After averaging, and due to the
particular structure of the operator ${\cal G}_0$, the mixed terms
will also give exponential decay when $t\to\infty$. They can thus be
discarded. Therefore, in the long time limit,
$\langle\delta\!c\rangle$ reads:
\begin{eqnarray}\fl
\langle\delta\!c\rangle
  = \lambda{\cal G}[\langle\delta\!\chi^2\rangle]
{}- 2\lambda(\sigma_{\rm\!A\!A}-\mu){\cal G}
    \Bigl[\langle\delta\!\chi{\cal G}[\delta\!\chi]\rangle\Bigr]
{}+ \lambda(\sigma_{\rm\!A\!A}-\mu)^2{\cal G}
    \Bigl[\langle{{\cal G}[\delta\!\chi]}^2\rangle\Bigr]
  + \cdots \label{eq:deltacapp}
\end{eqnarray}
where the ellipsis stands for terms containing at least fourth power
of $\delta\!\chi$. In the following, we shall show that they give
sub-leading contributions to the asymptotic time behavior of
$\langle\delta\!c\rangle$. Let us now analyze the first term of that
equation. It reads
\begin{eqnarray}
{\cal G}[\langle\delta\!\chi^2\rangle]
  = \int_0^t\d t'\!\int\!\d^dx\,\e^{-\sigma_{\rm\!A\!A}(t-t')}
    G_0({\bf x}-{\bf x}',t-t')\langle\delta\!\chi({\bf x}',t')^2\rangle
\end{eqnarray}
The large time behavior of this expression can be obtained in several
ways. One would be to integrate over the space dependence, and then to
use the property
\begin{equation}
\int_0^t\d t'\,\e^{-\alpha(t-t')}f(t')
  = \frac{1}{\alpha}f(t) + \Or(f'(t)),\quad (t\to\infty)
\label{eq:largetime}
\end{equation}
which of course is only valid if $f'(t)$ is negligible with respect to
$f(t)$ (in particular this property is false when $f(t)$ is an
exponential). In our case, one could safely use it, as
$\langle\delta\!\chi^2\rangle$ (which plays the role of $f(t)$)
decays with a power law. Thus one finds
\begin{equation}
{\cal G}[\langle\delta\!\chi^2\rangle]
  = \frac{1}{\sigma_{\rm\!A\!A}}
    \langle\delta\!\chi^2\rangle.
\end{equation}
Another equivalent way to obtain this result is to note that the
leading behavior of the time integral is obtained when $t'\to t$,
i.e.\ one can simply replace $\e^{-\sigma_{\rm\!A\!A}(t-t')}$ with
the delta function $\delta[\sigma_{\rm\!A\!A}(t-t')]$ and then use that
\begin{equation}
G_0({\bf x}-{\bf x}',0)=\delta^{(d)}({\bf x}-{\bf x}').
\end{equation}
In other words, in the long time limit, the operator $\cal G$ can
simply be replaced by $\sigma_{\rm\!A\!A}^{-1}$. Applying this method
for all the other terms of \eref{eq:deltacapp}, one readily obtains
the large time behavior for $\langle\delta\!c\rangle$:
\begin{eqnarray}
\langle\delta\!c\rangle
  = \biggl(\frac{\lambda}{\sigma_{\rm\!A\!A}}
  -        \frac{2\lambda(\sigma_{\rm\!A\!A}-\mu)}
                {\sigma_{\rm\!A\!A}^2}
  +        \frac{\lambda(\sigma_{\rm\!A\!A}-\mu)^2}
                {\sigma_{\rm\!A\!A}^3}
    \biggr)\langle\delta\!\chi^2\rangle
  + \cdots.
\end{eqnarray}
Note that the same result would have been obtained using the property
\eref{eq:largetime}. The terms containing higher powers of ${\cal
G}[\delta\!\chi]$ can be treated in the same way. They will give
(subleading) contributions of order $\langle\delta\!\chi^2\rangle^n$,
with $n\geq2$. Finally, one finds
\begin{eqnarray}
\langle\delta\!c\rangle
  = \frac{\lambda\mu^2}{\sigma_{\rm\!A\!A}^3}
    \langle\delta\!\chi^2\rangle
  + \cdots
  = \frac{2\lambda\mu^2}{\sigma_{\rm\!A\!A}^3}(c_0-c_\infty)
    (8\pi Dt)^{-d/2} + \cdots \label{eq:deltacfin}
\end{eqnarray}
Sub-leading corrections to that result are of order $t^{-d}$ or
$t^{-d/2-1}$, and will generally depend on microscopic details such as
the lattice constant, etc.

In summary, using the fact that $\delta\!\chi$ is a Gaussian variable
with a vanishing variance, we have written $\delta\!c$ as a power
series in $\delta\!\chi$, in a systematic way (formally, using this
method, one can compute the sub-leading corrections, however the
calculation may become rather tricky). The same method may be used to
obtain $\langle\delta\!c^2\rangle$, yielding
\begin{eqnarray}
\langle\delta\!c^2\rangle
& = \biggl(\frac{\sigma_{\rm\!A\!A}-\mu}{2\sigma_{\rm\!A\!A}}
    \biggr)^2\langle\delta\!\chi^2\rangle
  + \cdots, \nonumber \\
& = \frac12\biggl(\frac{\sigma_{\rm\!A\!A}-\mu}
                       {\sigma_{\rm\!A\!A}}\biggr)^2(c_0-c_\infty)
    (8\pi Dt)^{-d/2} + \cdots \label{eq:deltac2}
\end{eqnarray}

\Eref{eq:deltac2} implies that $\langle\delta\!c\rangle$ is not a
self-averaging quantity. Indeed, the relative density fluctuations are
given by
\begin{equation}
\Biggl|\frac{\langle\delta\!c^2\rangle - \langle\delta\!c\rangle^2}
            {\langle\delta\!c\rangle^2}
\Biggr|^{1/2}
  = \frac{1}{\sqrt2}\frac{\sigma_{\rm\!A\!A}-\mu}{\lambda\mu^2}
    \sigma_{\rm\!A\!A}^2{|c_0-c_\infty|}^{-1/2}(8\pi Dt)^{d/4}.
\end{equation}
They diverge in the long time limit. As a consequence, it is very
difficult to confirm numerically the validity of our result
\eref{eq:deltacfin}.
Indeed, denote by $\langle\Delta c\rangle$ the average
density measured in a typical simulation:
\begin{equation}
\langle\Delta c\rangle
\equiv \biggl\langle\frac{1}{L^d}\int\!\d^dx\,\delta\!c\biggr\rangle
  = \frac{2\lambda\mu^2}{\sigma_{\rm\!A\!A}^3}
    (c_0-c_\infty)(8\pi Dt)^{-d/2}
\end{equation}
(where $L$ is the system size). Its fluctuations are
\begin{eqnarray}
\langle\Delta c^2\rangle
&\equiv
    \biggl\langle\frac{1}{L^{2d}}\int\!\d^dx\!\int\!\d^dx'\,
    \delta\!c({\bf x},t)\delta\!c({\bf x}',t)\biggl\rangle
  = \frac{1}{L^d}\int\!\d^dx\,\langle\delta\!c(0,t)\delta\!c({\bf x},t)
    \rangle \nonumber \\
&\simeq
    \frac{1}{2L^d}\,
    \biggl(\frac{\sigma_{\rm\!A\!A}-\mu}
                {\sigma_{\rm\!A\!A}}\biggr)^2(c_0-c_\infty),
\end{eqnarray}
so that the relative fluctuations reads
\begin{equation}
\Biggl|\frac{\langle\Delta c^2\rangle-{\langle\Delta c\rangle}^2}
            {{\langle\Delta c\rangle}^2}\Biggl|^{1/2}
  = \frac{(\sigma_{\rm\!A\!A}-\mu)^2}{2\lambda\mu^2}\,\sigma_{\rm\!A\!A}^2\,
    |a_\infty-a_0|^{-1/2}(8\pi Dt/L)^{d/2}.
\end{equation}
The fluctuations are negligible only if $t\ll L$ (in contrast with
the diffusive behavior for which $t\ll L^2$), a regime which is
very difficult to obtain numerically.

Intriguingly, the same results for the density and its correlations
could be obtained by imposing $(\partial_t - D\nabla^2)\,\delta\!c =
0$ in equation \eref{eq:deltac} and then solving the quadratic
equation for $\delta\!c$. However this approximation is uncontrolled,
and only the lowest order terms can be obtained. It is possible to
explain why such a crude approximation works, by noting that even
though $\delta\!c$ is a random variable, it should not vary too
rapidly, as it only depends on $\delta\!\chi$ and not directly on the
noise $\zeta$.

Remark that the sign of $\langle\delta\!c\rangle$ is given by
$\langle\delta\!\chi^2\rangle$ (or equivalently by $c_0-c_\infty$). This
means that if initially $\langle\delta\!c\rangle$ is positive, so
will it be for large time. No information is given for intermediate
times, but a non-monotonic behavior would be surprising.

The computation of the density correlation functions goes along the
same lines. Using that $a=a_\infty + \delta\!\chi - 2\delta\!c$ and
equation \eref{eq:corrfct}, one obtains (when $t\to\infty$)
\begin{equation}\fl
C_a({\bf x},t)
  = a_\infty^2 + \delta^{(d)}({\bf x})
  + \frac{\mu^2}{\sigma_{\rm\!A\!A}^3}(a_\infty-a_0)(8\pi Dt)^{-d/2}
    \Bigl(\mu\e^{-x^2/8Dt} - 2\lambda\delta^{(d)}({\bf x})\Bigr),
\end{equation}
where the Gaussian factor $\e^{-x^2/8Dt}$ comes from the expression
for $\langle\delta\!\chi({\bf x},t)\delta\!\chi(0,t)\rangle$ (which is
easily obtained from \eref{eq:solchi}). The correlation length $\xi_a$
is given by
\begin{equation}
\xi_a^2
  = \biggl|\frac{\int\d^dx\,x^2 [C_a({\bf x},t)-a_\infty^2]}
                {\int\d^dx\,[C_a({\bf x},t)-a_\infty^2]}\biggr|
  = \frac{4d\pi Dt\,|a_\infty - a_0|}
                 {a_\infty - a_0 + a_\infty\sigma_{\rm\!A\!A}^3/\mu^3}
\end{equation}
The absolute values ensure the positiveness of $\xi_a^2$ (note that
$a_\infty - a_0 + a_\infty\sigma_{\rm\!A\!A}^3/\mu^3\geq0$). When
$a_0>a_\infty$, the second moment of $C_a({\bf x},t)-a_\infty^2$ is
negative, indicating that the $A$ particles are negatively
correlated. The same conclusion holds for the $C$ particles.

\section{The two-species reversible reaction
$A+B\protect\rightleftharpoons C$}\label{sec:ABC}

\subsection{The model}

In this section we study the reversible $A+B\rightleftharpoons C$
reaction using the previous approach. As before, our starting point is
the continuous time master equation describing the process on a
$d$-dimensional hypercubic lattice, allowing multiple occupancy. Let
$P(\{l\},\{m\},\{n\};t)$ be the probability to find the configuration
$\{l\}$, $\{m\}$, $\{n\}$ at time $t$. The set $\{l\}$ describe the
occupation numbers on each lattice site for the $A$ particles, $\{m\}$
is used for the $B$ particles and $\{n\}$ for the $C$'s. The master
equation is then
\begin{eqnarray}\fl
\frac{\partial}{\partial t} P(\{l\},\{m\},\{n\};t) \nonumber \\
\lo=\frac{D_a}{\ell^2}\sum_i\sum_{e_i}
    [(l_{e_i}+1)P(\ldots,l_i-1,l_{e_i}+1,\ldots,\{m\},\{n\};t)
    \nonumber \\
{}-  l_iP(\{l\},\{m\},\{n\};t)] \nonumber \\
{}+ \frac{D_b}{\ell^2}\sum_i\sum_{e_i}
    [(m_{e_i}+1)P(\{l\},\ldots,m_i-1,m_{e_i}+1,\ldots,\{n\};t)
    \nonumber \\
{}-  m_iP(\{l\},\{m\},\{n\};t)] \nonumber \\
{}+ \frac{D_c}{\ell^2}\sum_i\sum_{e_i}
    [(n_{e_i}+1)P(\{l\},\{m\},\ldots,n_i-1,n_{e_i}+1,\ldots;t)
    \nonumber \\
{}-  n_iP(\{l\},\{m\},\{n\};t)] \nonumber \\
{}+ \lambda_0\sum_i
    [(l_i+1)(m_i+1)P(\ldots,l_i+1,\ldots,\ldots,m_i+1,\ldots,\{n\};t)
    \nonumber \\
{}-  m_i(m_i-1)P(\{l\},\{m\},\{n\};t)] \nonumber \\
{}+ \mu\sum_i[(n_i+1)P(\{l\},\{m\},\ldots,n_i+2,\ldots;t)
    \nonumber \\
{}-           n_iP(\{l\},\{m\},\{n\};t)].
\label{eq:MEb}
\end{eqnarray}
This master equation has the same structure than equation
\eref{eq:ME}, and the same notation is used ($D_b$ is the diffusion
constant of $B$ particles). For the time being, we choose homogeneous
initial conditions given by an uncorrelated Poissonian distribution on
each site and for each species:
\begin{equation}
P(\{l\},\{m\},\{n\};0)
  = \e^{-\tilde a_0 - \tilde b_0 - \tilde c_0}
    \prod_i \frac{\tilde a_0^{l_i}}{l_i!}
    \frac{\tilde b_0^{m_i}}{m_i!}\frac{\tilde c_0^{n_i}}{n_i!},
\label{eq:CIb}
\end{equation}
where $\tilde b_0$ is the initial average occupation number of $B$
particles.

In the field theory formalism, the action is given by
\begin{eqnarray}\fl
S = \int\!\d^dx\!\int_0^{t_{\rm f}}\d t\,
    [\bar a(\partial_t - D_a\nabla^2)a + \bar b(\partial_t - D_b\nabla^2)b
  +  \bar c(\partial_t - D_c\nabla^2)c \nonumber \\
{}+  (\bar a\bar b + \bar a + \bar b - \bar c)(\lambda ab - \mu c)
  -  \delta(t)(a_0\bar a + b_0\bar b + c_0\bar c)] \label{eq:actb},
\end{eqnarray}
where $b_0=\tilde b_0/\ell^d$. In term of Langevin equations we find
\begin{eqnarray}
(\partial_t - D_a\nabla^2) a({\bf x},t)
  = -\lambda a({\bf x},t)b({\bf x},t) + \mu c({\bf x},t)
  + \zeta_a({\bf x},t) \label{eq:ALanb} \\
(\partial_t - D_b\nabla^2) b({\bf x},t)
  = -\lambda a({\bf x},t)b({\bf x},t) + \mu c({\bf x},t)
  + \zeta_b({\bf x},t) \label{eq:BLanb} \\
(\partial_t - D_c\nabla^2) c({\bf x},t)
  = \lambda a({\bf x},t)b({\bf x},t) - \mu c({\bf x},t), \label{eq:CLanb}
\end{eqnarray}
where $\zeta_a$ and $\zeta_b$ are two complex Gaussian noises with
zero mean value and whose correlation are given by
\begin{eqnarray}
\langle \zeta_a({\bf x},t)\zeta_a({\bf x}',t') \rangle
& = \langle \zeta_b({\bf x},t)\zeta_b({\bf x}',t') \rangle
  = 0 \\
\langle \zeta_a({\bf x},t)\zeta_b({\bf x}',t') \rangle
& = 2\langle \mu c({\bf x},t) - \lambda a({\bf x},t)b({\bf x},t) \rangle
    \delta^{(d)}({\bf x}-{\bf x}') \delta(t-t') \nonumber \\
& = -2\partial_t\langle c(t)\rangle
    \delta^{(d)}({\bf x}-{\bf x}') \delta(t-t')
\end{eqnarray}
(for homogeneous initial conditions). As for the
$A+A\rightleftharpoons C$ reaction, the noise vanishes at equilibrium.
The equilibrium densities are thus given by their mean-field solution:
\begin{equation}
\lambda a_\infty b_\infty = \mu c_\infty, \label{eq:steadyb}
\end{equation}
which together with the two conservation laws
\begin{eqnarray}
a_\infty + b_\infty + 2c_\infty = a_0 + b_0 + 2c_0 \\
a_\infty - b_\infty = a_0 - b_0,
\end{eqnarray}
permits us to obtain the equilibrium densities as a function of the
initial conditions.

\subsection{Conserved quantities}

The approach toward equilibrium will be obtained through the same
steps as before. First we write the Langevin equation for the conserved
quantities and then, once these equations solved, we can set up a
systematic approximation scheme for $\delta\!c$.

Let us introduce $\psi = a-b$ and $\chi = a+b+2c$. From equation
\eref{eq:ALanb}--\eref{eq:CLanb}, one readily obtains the two following
Langevin equations:
\begin{eqnarray}
(\partial_t-D\nabla^2)\psi({\bf x},t) = \zeta_\psi({\bf x},t), \\
(\partial_t-D\nabla^2)\chi({\bf x},t) = \zeta_\chi({\bf x},t),
\end{eqnarray}
where in order to simplify, we supposed $D_a=D_b=D_c\equiv D$ (the
case of different diffusion constants will be considered in
\sref{sec:ext}). The noises $\zeta_\psi$ and $\zeta_\chi$ have a
vanishing mean and their two-point correlations are
\begin{eqnarray}
\langle\zeta_\psi({\bf x},t)\zeta_\psi({\bf x}',t')\rangle
& = -\langle\zeta_\chi({\bf x},t)\zeta_\chi({\bf x}',t')\rangle
\nonumber \\
& = 2\partial_t\langle c(t)\rangle
    \delta^{(d)}({\bf x}-{\bf x}')\delta(t-t'). \\
\langle\zeta_\psi({\bf x},t)\zeta_\chi({\bf x}',t')\rangle
& = 0
\end{eqnarray}
The solution of these two Langevin equations has the same form as
equation \eref{eq:solchi}, with appropriate initial conditions
($\psi({\bf x},0) = a_0-b_0$ and $\chi({\bf x},0) =
a_0+b_0+2c_0$). $\psi$ and $\chi$ have then a Gaussian distribution
with a non vanishing mean
\begin{eqnarray}
\langle\psi\rangle = a_0-b_0, \qquad
\langle\chi\rangle = a_0+b_0+2c_0,
\end{eqnarray}
and their variance is
\begin{eqnarray}
\langle\psi^2\rangle - \langle\psi\rangle^2
  = -(\langle\chi^2\rangle - \langle\chi\rangle^2)
  = 2\int_0^t\d t_1{[8\pi D(t-t_1)]}^{-d/2}
    \partial_t\langle c(t_1)\rangle. \label{eq:psipsi}
\end{eqnarray}
In the long time limit, we find again the power law decay:
\begin{equation}
\langle\psi^2\rangle - \langle\psi\rangle^2
  = 2(c_\infty-c_0){(8\pi Dt)}^{-d/2},\qquad (t\to\infty).
\label{eq:psipsilong}
\end{equation}

\subsection{Approximation scheme}

The next step is to set up our approximation scheme for
$\langle\delta\!c\rangle\equiv c - c_\infty$. Let us define
$\delta\!\psi=\psi - (a_0-b_0)$ and $\delta\!\chi=\chi -
(a_0+b_0+2c_0)$. The Langevin equation for $\delta\!c$ reads
\begin{eqnarray}\fl
(\partial_t - D\nabla^2 + \sigma_{\rm\!A\!B})
\,\delta\!c({\bf x},t) \nonumber \\
\lo=\lambda\,\delta\!c({\bf x},t)^2
  - \lambda\,\delta\!\chi({\bf x},t)\,\delta\!c({\bf x},t)
  + \case14\lambda[\delta\!\chi({\bf x},t)^2 - \delta\!\psi({\bf x},t)^2]
    \nonumber \\
  + \case12(\sigma_{\rm\!A\!B}-\mu)\,\delta\!\chi({\bf x},t)
  - \case12\lambda(a_0-b_0)\,\delta\!\psi({\bf x},t),
\label{eq:deltacb}
\end{eqnarray}
where we put
$\sigma_{\rm\!A\!B}=\lambda(a_0+b_0+2c_0-2c_\infty)+\mu$. This
equation possesses the same structure as equation
\eref{eq:deltac}. By repeating the same scheme, we may then obtain
the large time behavior of $\langle\delta\!c\rangle$. Finally, we find
\begin{eqnarray}
\langle\delta\!c\rangle
  = \frac{\lambda\mu^2}{4\sigma_{\rm\!A\!B}^3}
    \langle\delta\!\chi^2\rangle
  - \frac{\lambda}{4\sigma_{\rm\!A\!B}}
    \Biggl[1
  -       \biggl(\lambda\,\frac{a_0-b_0}{4\sigma_{\rm\!A\!B}}\biggr)^2
    \Biggr]\langle\delta\!\psi^2\rangle
  + \cdots
\end{eqnarray}
or, by putting the large time expression for
$\langle\delta\!\chi^2\rangle$ and $\langle\delta\!\psi^2\rangle$
\begin{eqnarray}
\langle\delta\!c\rangle
  = \frac{\lambda}{2\sigma_{\rm\!A\!B}^3}
    [\sigma_{\rm\!A\!B}^2 + \mu^2 - \lambda^2(a_0-b_0)^2]
    (c_0-c_\infty)(8\pi Dt)^{-d/2} + \cdots
\end{eqnarray}
Note that $\sigma_{\rm\!A\!B}^2 + \mu^2 - \lambda^2(a_0-b_0)^2 =
2\mu(\sigma_{\rm\!A\!B} + 2\lambda c_\infty) > 0$.

Similarly, one obtains the $C$-particle correlation:
\begin{equation}
\langle\delta\!c^2\rangle
  = \frac{2\lambda\mu c_\infty}{\sigma_{\rm\!A\!B}^2}
    (c_0-c_\infty)(8\pi Dt)^{-d/2} + \cdots \label{eq:deltacbfin}
\end{equation}
Sub-leading corrections to these laws are of order $t^{-d}$ or
$t^{-d/2-1}$.

\section{Extensions}
\label{sec:ext}

\subsection{Correlations in the initial condition}

Recently, Yang \etal \cite{YLS} showed that in the
$A+B\rightleftharpoons C$ case, and for correlated initial conditions,
the power law approach \eref{eq:deltacbfin} could be modified. More
precisely, for a particular initial condition, they found that in one
and three dimensions the approach to equilibrium was of the order
$t^{-d/2-1}$, i.e.\ faster than in the uncorrelated case. Whereas it
seems to be at odds with the fact that the power law should be
universal (see the final section), we shall show in this subsection
how such a behavior can be obtained within our approach.

Let us consider the reaction $A+A\rightleftharpoons C$, when initially a
fraction of the total density of the $A$ particles are not distributed
independently, but are disposed in pairs having a separation radius
$\sigma$ (each $A$-$A$ pairs are supposed to be independently
distributed). We still denote the total density by $a_0$, and the
density of pairs will be denoted by $n_0$. In order to take this
condition into account in our formalism, one should add the following
term in the action \eref{eq:act}
\begin{equation}
-\frac{n_0}{s_d\sigma^{d-1}}\int\!\d^dx\!\int\!\d^dy\,
\bar a({\bf x},0)\bar a({\bf y},0)\delta(|{\bf x} - {\bf y}| - \sigma).
\end{equation}
($s_d$ is the surface of $d$ dimensional sphere of radius unity). In
the language of Langevin equations, this new term translates into a new
contribution to the noise correlation, namely
\begin{equation}\fl
\langle \zeta({\bf x},t)\zeta({\bf x}',t') \rangle
  = -2\partial\langle c(t)\rangle\delta^{(d)}({\bf x}-{\bf x}')\delta(t-t')
  + \frac{n_0}{s_d\sigma^{d-1}}\,2\delta(|{\bf x} - {\bf x}'|-\sigma)
    \delta(t)\delta(t').
\end{equation}
The only relevant effect of this new contribution is to affect the
variance of $\chi$, to which one should add
\begin{eqnarray}
\langle\delta\!\chi^2\rangle_{\rm corr}
& = \frac{2n_0}{s_d\sigma^{d-1}}\int\!\d^dx'\!\int\d^dx''\,
    G_0({\bf x}-{\bf x}',t)G_0({\bf x}-{\bf x}'',t)
    \delta(|{\bf x} - {\bf x}'|-\sigma) \nonumber \\
& = 2n_0\exp(-3\sigma^2/4Dt)(8\pi Dt)^{-d/2}. \label{eq:chichicor}
\end{eqnarray}
For long time, $\langle\delta\!\chi^2\rangle$ becomes then (we used
that $2(c_0-c_\infty)=a_\infty-a_0$)
\begin{equation}
\langle\delta\!\chi^2\rangle
  = [a_\infty-(a_0-2n_0)](8\pi Dt)^{-d/2} + \cdots
\end{equation}
$a_0-2n_0$ represents the initial density of uncorrelated $A$
particles. Inserting this result into formula \eref{eq:deltacfin} one obtains
the $C$-particle density:
\begin{eqnarray}
\langle\delta\!c\rangle
  = \frac{\lambda\mu^2}{\sigma_{\rm\!A\!A}^3}(a_\infty-a_0+2n_0)
    (8\pi Dt)^{-d/2} + \cdots
\end{eqnarray}
Initial correlations could then lead to non-monotonic behavior
of $\langle\delta\!c\rangle$, as $a_\infty-a_0$ and
$a_\infty-a_0+2n_0$ do not have necessarily the same sign.

A case of special interest occurs when we choose $2n_0$ to be exactly
$a_\infty-a_0$. The amplitude of the leading term of the approach to
equilibrium vanishes and one should consider the sub-leading corrections
to $\langle\delta\!\chi^2\rangle$. As a consequence the approach to
equilibrium will be faster. It is easy to verify that these corrections
will be of order $t^{-d/2-1}$, with an amplitude that depends on
microscopic parameters such as $\sigma$, the initial correlation
length, or $\ell$, the lattice constant. Note that the condition
$2n_0=a_\infty-a_0$ can be easily obtained experimentally: consider a
system where $A$ and $C$ particles are at equilibrium. At time $t=0$, we
excite some fraction of $C$ particles such that they break up into pairs
of correlated $A$ particles (this excitation can be for example
obtained by a photo-flash \cite{YLS}). Let us denote by $\tilde
a_\infty$ and $\tilde c_\infty$ the concentrations of $A$ and $C$
particles before the excitation (note that we were at equilibrium:
$\lambda \tilde a_\infty^2 = \mu \tilde c_\infty$). At time $t=0$, the
initial densities will be $a_0=\tilde a_\infty + 2 n_0$ and $c_0 =
\tilde c_\infty - n_0$. Now, as $a_0+2c_0=\tilde a_\infty + 2\tilde
c_\infty$, one must then have $a_\infty=\tilde a_\infty$, which
implies $2n_0=a_0-a_\infty$. This initial condition correspond exactly
to the case studied by Yang \etal \cite{YLS}, for the two-species
reaction.

Similar conclusions can be drawn for the $A+B\rightleftharpoons C$
reaction when initially a given amount of $A$ and $B$ particles are
distributed in pairs of radius $\sigma$ (initial correlations among
particles of the same species will not be considered, as they should
decay exponentially fast). The action get modified in the same way as
previously, by adding a new initial term:
\begin{equation}
-\frac{n_0}{s_d\sigma^{d-1}}\int\!\d^dx\!\int\!\d^dy\,
\bar a({\bf x},0)\bar b({\bf y},0)\delta(|{\bf x} - {\bf y}| - \sigma).
\end{equation}
In terms of the Langevin equation it also modifies the noise $\zeta$,
leading to a new contribution for both $\langle\delta\!\psi^2\rangle$
and $\langle\delta\!\chi^2\rangle$ (see \eref{eq:chichicor}). Finally,
one gets
\begin{equation}
\langle\delta\!\chi^2\rangle = -\langle\delta\!\psi^2\rangle
  = [a_\infty-(a_0-n_0)](8\pi Dt)^{-d/2} + \cdots
\end{equation}
Here $n_0$ is the initial concentration of correlated pairs, it is
also the initial density of correlated $A$ (or $B$)
particles. Contrarily to the irreversible $A+B\to\emptyset$ case, the
presence of correlations in the initial state does not modify the long
time behavior (except in some special limits as discussed above).

\subsection{Unequal diffusion coefficient}

The case of unequal diffusion constant is of special interest, because
if, for some reasons, one kind of particles moves very slowly compares
to the other, one could legitimately question the validity of our
previous results. In fact we shall see that even in the worse case
(one species at rest) the power law does not change, only the
amplitude is affected.

Let us consider first the $A+A\rightleftharpoons C$ reaction, the
generalization to the $A+B\rightleftharpoons C$ reaction being
straightforward. The two Langevin equations associated with that model are
\begin{eqnarray}
(\partial_t - D_a\nabla^2)\,\delta\!\chi
  + 2(D_a-D_c)\nabla^2\,\delta\!c = \zeta, \label{eq:Lanchi} \\
(\partial_t - D_c\nabla^2 + \sigma_{\rm\!A\!A})\,\delta\!c
  = 4\lambda\,\delta\!c^2 - 4\lambda\,\delta\!\chi\,\delta\!c
  + \lambda\,\delta\!\chi^2
  + \case12(\sigma_{\rm\!A\!A}-\mu)\,\delta\!\chi \label{eq:Lanc}
\end{eqnarray}
The first consequence of having two different diffusion constants lies
in the fact that the Langevin equation for $\delta\!\chi$ is no longer
closed, but contains a term proportionnal to $\delta\!c$. The main
question we want to address, is how does this extra term affect the
large time behavior of the $\langle\delta\!\chi^2\rangle$? One quick
and false answer would be to say that, as $\delta\!c$ is a slowly
varying random variable for large time, it should not modify the large
time behavior of $\delta\!\chi$. This picture is however not true, as
$\delta\!c$ itself depends on $\delta\!\chi$ in a non trivial way. In
fact, the better way to treat this term is to insert it into the
propagator, which is then no longer diagonal.
Putting also in it the term proportionnal to $\delta\!\chi$ in
equation \eref{eq:Lanc} (which was previously treated like an
interaction term), the propagator is given, in Fourier and Laplace transform
representation, by the inverse of the following matrix
\begin{equation}
{\cal M}
  = \left(\begin{array}{cc}
          s + D_a p^2  &  -2 (D_a-D_c) p^2 \\
          -\case12(\sigma_{\rm\!A\!A}-\mu)
                       &    s + D_c p^2 + \sigma_{\rm\!A\!A}
          \end{array}
    \right)
\end{equation}
where $\bf p$ and $s$ are the Fourier and Laplace conjugated variables
of space and time (this last expression is very easily obtained in the
field theory formalism by considering the quadratic terms). For the
$\delta\!\chi\delta\!\chi$ propagator, one then finds in the $({\bf
p},s)$ representation
\begin{equation}
G_{\delta\!\chi\delta\!\chi,0}({\bf p},s)
  = \frac{s + D_cp^2 + \sigma_{\rm\!A\!A}}
         {(s+D_ap^2+\sigma_{\rm\!A\!A})(s+D_cp^2)+(D_a-D_c)\mu p^2}.
\end{equation}
The $\delta\!\chi\delta\!c$ propagator is given by
\begin{equation}
G_{\delta\!\chi\delta\!c,0}({\bf p},s)
  = \frac{2 (D_a-D_c) p^2}
         {(s+D_ap^2+\sigma_{\rm\!A\!A})(s+D_cp^2)+(D_a-D_c)\mu p^2}.
\label{eq:chicprop}
\end{equation}
The two other propagators could be obtained in the same way, but we
are not interested by them. Our purpose now is to compute the large
time limit of $\langle\delta\!\chi^2\rangle$. Once this obtained, we
shall then use our approximation scheme (applied to equation
\eref{eq:Lanc}), in order to derive its large time behavior.

\begin{figure}[tb]
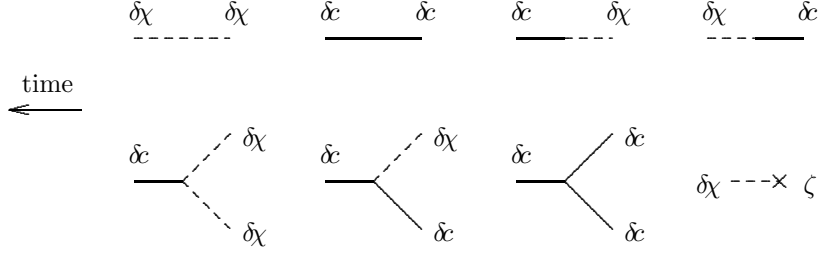

\centerline{\input reyfig1.tex}
\caption{Set of propagators (upper part) and vertices (lower part)
needed for the diagrammatic representation of the Langevin
equations. The noise term is represented beginning with a cross. The
arrow on the left represent the direction of the
time. \label{fig:vertices}}
\end{figure}

For computing the second moment of $\delta\!\chi$, it is easier to
consider a diagrammatic representation (in this paragraph, we follow
the line of reasoning developped in \cite{LC} for the irreversible
reaction $A+B\to\emptyset$, when $D_a\neq D_C$). A
$\delta\!\chi\delta\!\chi$ propagator will be represented by a dashed
line and a $\delta\!c\delta\!c$ propagator by a full
line. Off-diagonal propagators will be represented by mixing of the
two lines (see \fref{fig:vertices}). From the Langevin equation
\eref{eq:Lanc}, three different vertices (four with the noise) can be
identified, their representation is also given in
\fref{fig:vertices}. As we shall eventually average over the noise,
only diagrams containing two merging noise lines at their beginning
can subsist. To obtain the second moment of $\delta\!\chi$, we simply
have to draw all diagrams ending with two merging $\delta\!\chi$ lines
(see \fref{fig:chichi}). If $D_a=D_c$, only the first diagram would
give a non-vanishing contribution (in agreement with equation
\eref{eq:chichi}). In fact, the subsequent terms of the sum all
contain at least one of the three sub-diagrams shown in
\fref{fig:irrel}. However, in the language of field theory, these
sub-diagrams give rise to effective vertices of the form
$\bar{\delta\!\chi}\nabla^2\delta\!\chi^2$,
$\bar{\delta\!\chi}\nabla^2\delta\!\chi\delta\!c$ and
$\bar{\delta\!\chi}\nabla^2\delta\!c^2$, which, by simple power
counting, turn out to be irrelevant. They will only give sub-leading
contribution to $\langle\delta\!\chi^2\rangle$. Hence, its leading
term is only given by the first diagram of \fref{fig:chichi}.
\begin{figure}[tb]
\centerline{\input reyfig2.tex}
\caption{Diagrammatic representation of
$\langle\delta\!\chi^2\rangle$. The dot stands for noise-noise
correlation. \label{fig:chichi}}
\end{figure}
\begin{figure}[tb]
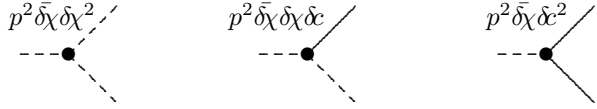

\centerline{\input reyfig3.tex}
\caption{Effectives vertices appearing in all diagrams besides the
first one of \protect\fref{fig:chichi}. They all give sub-leading
contribution to $\langle\delta\!\chi^2\rangle$. \label{fig:irrel}}
\end{figure}

To compute this contribution one first needs to obtain
$G_{\delta\!\chi\delta\!\chi,0}$ in $({\bf p},t)$ space. Inverting the
Laplace transform, one finds
\begin{eqnarray}\fl
G_{\delta\!\chi\delta\!\chi,0}({\bf p},t)
  = \frac{1}{r({\bf p})}
    \exp[-\case12(D_a+D_c)p^2 t - \case12\sigma_{\rm\!A\!A}t]
    \nonumber \\
{}\times
    \{[\sigma_{\rm\!A\!A} - (D_a-D_c)p^2]\sinh[\case12r({\bf p})t]
  +   r({\bf p})\cosh[\case12r({\bf p})t]\}\theta(t)
\end{eqnarray}
where
\begin{equation}
r({\bf p}) = \Bigl({(\sigma_{\rm\!A\!A} + (D_a-D_c)p^2)}^2
     -       4\mu(D_a-D_c)p^2\Bigr)^{1/2}.
\end{equation}
One is now in position to obtain $\langle\delta\!\chi^2\rangle$. To
leading order, it is given by
\begin{eqnarray}
\langle\delta\!\chi^2\rangle
  = -2\int_0^t\d t'\!\int\!\frac{\d^dp}{(2\pi)^d}\,
    {[G_{\delta\!\chi\delta\!\chi,0}({\bf p},t-t')]}^2
    \partial_t\langle\delta\!c(t')\rangle,
\end{eqnarray}
whose large time behavior (which can be obtained by taking the large
$\sigma_{\rm \!\!A\!A}$ and $\mu$ limit) reads
\begin{equation}
\langle\delta\!\chi^2\rangle
  = -2 (8\pi D_{\rm eff})^{-d/2}\int_0^t\d t'\,(t-t')^{-d/2}
    \partial_t\langle\delta\!c(t')\rangle + \cdots
\end{equation}
with 
\begin{equation}
D_{\rm eff} = D_c + (D_a - D_c)\frac{\mu}{\sigma_{\rm\!A\!A}}
\label{eq:Deff}
\end{equation}
This is nothing but the result of equation \eref{eq:chichi}, with $D$
replaced by $D_{\rm eff}$.

We can now come back to the Langevin equation for $\delta\!c$ which
is exactly the same than the one written for equal diffusion constant
(equation \eref{eq:deltac}), with $D$ replaced by $D_c$, and then use
our approximation scheme as before. The final expression for
$\langle\delta\!c\rangle$ will then be given by equation
\eref{eq:deltacfin} with $D$ replaced by $D_{\rm eff}$.

It is particularly interesting to consider the cases when either $D_a$
or $D_c$ vanish. If $D_c=0$, one finds $D_{\rm eff}=D_a\mu/\sigma_{\rm
\!\!A\!A}$. Although the $C$ particles do not move, one still observes
a power law decay of the concentration, but with a smaller diffusion
constant than when $D_c=D_a$ ($\mu\leq\sigma_{\rm\!A\!A}$). The fact
that the $C$ particles are at rest is compensated by the movement of
the $A$ particles, which leads to an effective diffusion of the $C$
particles. More surprising is the case $D_a=0$. At first sight, one
could expect that because the $A$ particles do not move, the forward
reaction is essentially inoperative (only $A$ particles at the same
site could react), however, one should not forget that the $C$
particles still move and that they effectively carry two $A$ particles
(allowing thus some mixing of the $A$ particles), and secondly this
motion still allows the fluctuations of $\delta\!\chi$ to be smoothened
diffusively.

For the $A+B\rightleftharpoons C$ reaction, the case where $D_c$
differs from $D_a=D_b$ can be studied in exactly the same way. All we
said for the $A+A\rightleftharpoons C$ reaction is still valid, with
the modification that, as $\delta\!\psi$ only depends on the random
variables $a$ and $b$, it is not modified. Only $\delta\!\chi$
changes. As a consequence, the formula \eref{eq:psipsilong} is still
valid, and for $\delta\!\chi$ one finds
\begin{equation}
\langle\delta\!\chi^2\rangle
  = 2(c_\infty-c_0){(8\pi D_{\rm eff}t)}^{-d/2},\qquad (t\to\infty),
\end{equation}
where $D_{\rm eff}$ is given by equation \eref{eq:Deff} with
$\sigma_{\rm\!A\!A}$ replaced by $\sigma_{\rm\!A\!B}$. The expression
for $\langle\delta\!c\rangle$ becomes then
\begin{eqnarray}\fl
\langle\delta\!c\rangle
  = \frac{\lambda}{2\sigma_{\rm\!A\!B}^3}
    \biggl[\mu^2
          \biggl(1-\frac{\sigma_{\rm\!A\!B}-\mu}{\sigma_{\rm\!A\!B}}
                \frac{D_a-D_c}{D_a}\biggr)^{-d/2}
  +       \sigma_{\rm\!A\!B}^2 - \lambda^2(a_0-b_0)^2
    \biggr] \nonumber \\
{}\times
    (c_0-c_\infty)(8\pi D_at)^{-d/2}.
\end{eqnarray}
When $D_c=0$, the amplitude is slightly modified (compared to the case
$D_c=D_a$) but one could still consider the diffusion constant to be
$D_a$. Note that the case $D_a=0$ cannot be treated by this formalism.

The case where all the diffusion constants are different can be
considered as well. We shall not treat it here, but the result
should not differ too much from the previous one. Indeed the main new
ingredient is that $\langle\delta\!\psi^2\rangle$ is modified. However
it has been shown that, for the $A+B\to\emptyset$ reaction, this leads
only to a change in the amplitude \cite{LC}. We expect this result to
be only slightly modified in the reversible case. For $\delta\!\chi$
the same analysis as before applies, but with more complicated
expressions. The cases $D_a=0$ or $D_b=0$ cannot be treated with this
formalism.

\subsection{Segregated initial conditions}

Our approach can be extended as well to other initial conditions. The
$A+B\rightleftharpoons C$ reaction with initially segregated reactants
(say, the $A$ particles on the right, the $B$ ones on the left and no
$C$ particles) is of particular interest, mainly due to the dynamics
of the reaction front. In the irreversible case, it is now well
established that the width of the front increase with a power law
$w(t)\sim t^{\alpha}$ \cite{GR,CD,LC2}. Two different cases may be
distinguished: above two dimensions, the exponent takes its mean-field
value $\alpha=1/6$, whereas below two dimensions fluctuation effects
play a dominant role, leading to $\alpha=1/[2(d+1)]$. Extrapolating our
previous results, one could expect that the fluctuations in the conserved
quantities will play an important role. In fact it appears that the
width of the front increases indeed with a power law $w(t)\sim
t^{\alpha}$, but this exponent takes its mean-field value
($\alpha=1/2$) for {\it any}\/ dimensions. Thess surprising results
have already been obtained by Chopard \etal \cite{CDKR} using scaling
arguments and numerical simulations. In this subsection, we shall show
how this behavior can be confirmed within our formalism.

The problem is described by the Langevin equations
\eref{eq:ALanb}--\eref{eq:CLanb} with the initial conditions
\begin{eqnarray}
\psi({\bf x},0) = n_0[\theta(x_1)-\theta(-x_1)] \\
\chi({\bf x},0) = n_0 \\
c({\bf x},0) = 0.
\end{eqnarray}
For simplicity we have chosen the particle to have the same diffusion
constants; moreover both reacting species $A$ and $B$ are supposed to
be homogeneously distributed (with density $n_0$) in their respective
semi-infinite sub-space. Let us denote by $\hat c$ the mean-field $C$
particles density. One easily shows that in the long time limit, it
takes a scaling behavior:
\begin{equation}
\hat c({\bf x},t) \simeq c_{\infty}(\xi),
\end{equation}
where $\xi=x_1/\sqrt{4Dt}$. The exact form of $c_{\infty}(\xi)$ is
unimportant. It can be obtained by equating to zero the right hand side
of the rate equation for the $C$ particles. Defining the width of the
front by the square root of the second moment the density of $C$
particles, one immediately obtains, in the mean-field case, the
exponent $\alpha=1/2$ for the reaction front.

In order to take into account the fluctuations, one needs first to
integrate the equation for the conserved quantities $\psi$ and
$\chi$. In the long time limit, one can show that
\begin{equation}
\langle\delta\!\psi^2\rangle
  = -\langle\delta\!\chi^2\rangle
  \simeq
  - (8\pi Dt)^{-d/2}\int_{-\infty}^{+\infty}\frac{\d y}{\sqrt\pi}\,
    \e^{-y^2}c_\infty(\xi-y).
\end{equation}
The Langevin equation for the $C$ particles can be rewritten using
$\delta\!c\equiv c - \hat c$, so that
\begin{eqnarray}\fl
[\partial_t - D\nabla^2]\delta\!c
  = \case14\lambda(\delta\!\chi^2 - \delta\!\psi^2)
  + \lambda\delta\!c^2 - \lambda\delta\!\chi\delta\!c \nonumber \\
{}+ \case12[\sigma(x_1,t)-\mu]\delta\!\chi
  - \case12\lambda n_0\,\hbox{erf}(\xi)\,\delta\!\psi
{}+ \sigma(x_1,t)\delta\!c
\end{eqnarray}
where $\sigma(x_1,t)=\lambda n_0 - 2\lambda \hat c(x_1,t) + \mu$, and
$\hbox{erf}$ is the error function. This equation is very similar to
equation \eref{eq:deltacb}, with the difference that $\sigma$ has
become time and position dependent. As a consequence, we are unable to
write in a closed form the propagator related to this equation, and
thus to apply our approximation scheme. However we can still deduce
the large time behavior of the density $\langle\delta\!c\rangle$ by
equating the right hand side to 0 and solving the quadratic equation
in $\delta\!c$. This crude approximation has proven to give accurate
results in the homogeneous one- and two-species reactions. In
addition, the same results are obtained by assuming (on physical
grounds and by examining its differential equation) that the
propagator behaves like
\begin{equation}
G({\bf x},{\bf x}',t,t')
\simeq \exp[-\sigma(x_1,t)(t-t')] G_0({\bf x}-{\bf x}',t-t')
\end{equation}
when $t'\to t$ (note that in the homogeneous case, this limit gave the
asymptotic time behavior).

Finally, one finds
\begin{equation}
\langle c \rangle
  = \hat c(x_1,t)
   + \frac{\lambda}{4{\sigma(x_1,t)}^3}
     \{\mu^2 + {\sigma(x_1,t)}^2 - \lambda^2n_0^2{[\hbox{erf}(\xi)]}^2\}
     \langle\delta\!\chi^2\rangle.
\end{equation}
As in the homogeneous case, the mean-field asymptotic solution is
approached with a power law. From the last equation one immediately
obtains that the width of the front is governed by its mean-field
exponent $\alpha=1/2$. This result is easily explained: the spreading
of the front is given by the diffusion of the $C$ particles. Moreover,
once the backward reaction is allowed, it has been shown that the $C$
particles will always diffuse with a non-vanishing effective diffusion
constant (even when $D_c=0$). Hence the width of the front should grow
like the square root of time, independently on the fluctuations which
are governing only the approach to the equilibrium, and not the
spreading of the $C$ particles.

\section{Discussion and concluding remarks}
\label{sec:conc}

As we mentioned in the beginning, our model allows for multiple
occupancy of each site and contains only single site reactions, a
property which has considerably simplified our analysis (leading in
particular to the simple form of the Langevin equations). It is thus
natural to question about the universality of our results.

To answer this question, let us consider the following two Langevin
equations
\begin{eqnarray}
(\partial_t - D\nabla) \Psi = \zeta \label{eq:consG}\\
(\partial_t - D\nabla) \Phi
  = \sum_{i,j\atop i+j\geq1}a_{i,j}\Psi^i\Phi^j, \label{eq:dens}
\end{eqnarray}
with $\langle\zeta\rangle=0$ and
\begin{equation}
\langle \zeta({\bf x},t)\zeta({\bf x}',t') \rangle
  = \Gamma(t)\delta^{(d)}({\bf x}-{\bf x}')\delta(t-t')
\end{equation}
(homogeneous case). By writing these two equations, we have
implicitely assumed that initially $\Psi(x,0)=0$ and that
$\lim_{t\to\infty}\langle\Phi\rangle=0$ (otherwise a constant term
$a_{00}\neq0$ should be added). From equation \eref{eq:consG}, one
immediately obtain in the long time limit
\begin{equation}
\langle \Psi^2 \rangle
  = (8\pi D)^{-d/2}\int_0^t\d t'\,(t-t')^{-d/2}\Gamma(t')
\end{equation}
(to cure the divergence when $t'\to t$, the integrand should be
multiplied by a cut-off function), leading to the following long time
behavior for $\langle\Phi\rangle$
\begin{equation}\fl
\langle \Phi \rangle
  = -\frac{1}{a_{01}^3}(a_{01}^2a_{20}-a_{01}a_{11}a_{10}+a_{02}a_{10}^2)
    (8\pi D)^{-d/2}\int_0^t\d t'\,(t-t')^{-d/2}\Gamma(t').
    \label{eq:gen}
\end{equation}
If $a_{01}\neq 0$, the leading behavior is given by $\int_0^t\d
t'\,(t-t')^{-d/2}\Gamma(t')$ (as long as
$a_{01}^2a_{20}-a_{01}a_{11}a_{10}+a_{02}a_{10}^2 \neq 0$). If
$\int_0^{\infty}\d t'\,\Gamma(t')$ is finite (this implies in
particular that the noise dies out at equilibrium), one recovers the
$t^{-d/2}$ power law. This shows us that in general the power law does
not depend on the structure of the equation for $\Phi$, but only on
the presence of $\Psi$, i.e.\ of a diffusive conserved mode in the
model. The condition $a_{01}\neq0$ (here $a_{01}$ plays the role of
$\sigma_{\rm\!A\!A}$ or $\sigma_{\rm\!A\!B}$) implies the presence of
a non-vanishing mass in the field theory for the field $\Phi$. This is
responsible for the presence of an exponential decay in time of the
$\Phi\Phi$ propagator, which, in turn, plays a central role in the
large time behavior.

\Eref{eq:gen} shows that the amplitude depends only on the most
relevant operators (the value of the coefficient $a_{i,j}$, with
$1\leq i+j\leq2$). In particular, a model with an exclusion principle
on each lattice site can be handled in the same way. Whereas the
exclusion condition will give rise to new contributions for the noise
correlation and for various operators, one can reasonably expect on
physical grounds that the conserved quantity will behave diffusively,
leading to the $t^{-d/2}$ power law. However, the equation for the
evolution of $\Phi$ will be modified, leading to new expressions for
the steady-state densities (which are again simply obtained by solving
the corresponding mean-field equation) and the amplitude. Note that
the exact expression of the equilibrium densities depends explicitly
on the complete equation for (the unshifted version of) $\Phi$, but as
the noise dies out at equilibrium, it is still given by a mean-field
equation.

The addition of a new diffusive conserved mode (like in the case of
the $A+B\rightleftharpoons C$ reaction) clearly does not alter the
power law, but the amplitude. A generalization of our analysis is also
straightforward for the reversible aggregation
$A_m+A_n\rightleftharpoons A_{m+n}$, for which the total number of
monomers $A$ is conserved. It is however not possible to extend our
results for the reversible coagulation process $A+A\rightleftharpoons
A$, for which no such conservation law occurs. In fact this last
reaction can be described by an action equivalent to the one obtained
for directed percolation, in the active phase. One expects thus the
upper critical dimension to be 4.

\ack
We thank Prof.\ Zolt\'an R\'acz, for many fruitful
discussions. P-A R is supported by the Swiss National Science
Foundation. This research was supported in part by EPSRC Grant
ER/J78327.

\end{document}